\documentclass[runningheads,a4paper]{llncs}
\usepackage[utf8]{inputenc}
\usepackage{enumitem}
\usepackage{graphicx}
\usepackage{eurosym}
\usepackage{float}
\usepackage{varwidth}
\usepackage{url}
\usepackage{amsmath}
\usepackage{amssymb}
\usepackage{caption}
\usepackage{subfigure}

\begin{document}

\title{Reciprocal Recommender System for Learners in Massive Open Online Courses (MOOCs)}
%
%
%
%
%

\author{Sankalp Prabhakar \inst{1} \and Gerasimos Spanakis \inst{2} \and Osmar Zaiane \inst{1}}

\institute{University of Alberta, Edmonton AB T6G 2R3, Canada\\
\email{\{sankalp, zaiane\}@ualberta.ca},
\and
Maastricht University, 
Maastricht, 6200MD, Netherlands\\
\email{jerry.spanakis@maastrichtuniversity.nl}}

\titlerunning{Reciprocal Recommendation for MOOCs}

\maketitle
\begin{abstract}
Massive open online courses (MOOC) describe platforms where users with completely different backgrounds subscribe to various courses on offer. MOOC forums and discussion boards offer learners a medium to communicate with each other and maximize their learning outcomes. However, oftentimes learners are hesitant to approach each other for different reasons (being shy, don't know the right match, etc.). In this paper, we propose a reciprocal recommender system which matches learners who are mutually interested in, and likely to communicate with each other based on their profile attributes like age, location, gender, qualification, interests, etc. We test our algorithm on data sampled using the publicly available MITx-Harvardx dataset and demonstrate that both attribute importance and reciprocity play an important role in forming the final recommendation list of learners. Our approach provides promising results for such a system to be implemented within an actual MOOC.
\end{abstract}


\section{Introduction}
Higher education is an area that has thus far embraced, but arguably has not been fundamentally altered by the growth of the Internet. This has been rapidly changing over the last few years with the rise of Massive Open Online Courses (MOOCs) as a way of learning that lets students participate on their own terms and conditions via Internet. Number of students that signed up for at least one course in year 2015 has crossed 35 million - up from an estimated 16-18 million the previous year \footnote{data collected by https://www.class-central.com}.

MOOC courses integrate the connectivity of social networks, the facilitation of an acknowledged expert in a field of study and a collection of freely accessible online resources. MOOC learners are diverse, originating from many cultures across the globe in all ages and backgrounds \cite{ho2015harvardx}. Despite this diversity, three main attributes unite them: A desire to learn, a desire to connect to a global community and a desire to experience and consume content online. Our work focuses on exploring the possibilities of assisting MOOC learners in the process of self-organization (e.g. forming study groups, finding partners, encourage peer learning, etc.) by developing a reciprocal recommender system that will recommend learners to each other based on a predefined set of preferences (e.g. interests, age range, location, qualification, gender, etc.). Moreover, lack of effective student engagement is one of main reasons for a very high MOOC dropout rate \cite{onah2014dropout}. Although many thousands of participants enroll in various MOOC courses, the completion rate for most courses is below 13\%. Further studies \cite{freeman2014active}, \cite{zepke2010improving} have been made to show how collaboration or active learning promotes student engagement. Therefore, we believe that recommending learners to each other will foster better student collaboration and would help mitigate the dropout rates to some extent.

The remainder of the paper is organized as follows. Section 2 presents some related work on the criteria of recommendations in MOOCs. In Section 3, we talk about the data and the proposed model for generating and ranking recommendations. Soon after, in Section 4, experimental results and evaluation are presented. Lastly, Section 5 concludes the paper and presents future work.

\section{Related Work}

Recommender systems for MOOCs have been developed, but their main focus has been on recommending courses to learners  \cite{apaza2014online}, \cite{bousbahi2015mooc}. i-Help peer recommendation \cite{Bull:2001:UMI:647664.733416} was an early effort towards this area but the matchmaking process is not clear and there is no evaluation of the results. People-to-people recommender systems have been studied in the general context \cite{koprinska2015people} involving techniques such as collaborative filtering, semantic-based methods, data mining, and context-aware methods as well as testing performance and effect of recommender systems. However, they have not found much application in the context of education.

Some of the most significant work in reciprocal recommendation has been done in the domain of online dating. The subject is more relevant here because a successful match only occurs when both recommended people like each other or reciprocate. In their work \cite{akehurst2011ccr}, authors built a Content-Collaborative Reciprocal (CCR) system. The content-based part uses selected user profile features and similarity measure to generate a set of similar users. The collaborative filtering part uses the interactions of the similar users, including the people they like/dislike and are liked/disliked by, to produce reciprocal recommendations. Other approaches include RECON \cite{pizzato2010recon}, a reciprocal recommender system for online dating which utilizes user preferences to calculate compatibility scores for each other.

Our research draws inspiration from some of the works mentioned above. More specifically, our system takes into account one of the MOOC particularities: there is no extended history for learners' preferences, thus traditional collaborative filtering systems are not directly applicable. Moreover, the idea of reciprocity and peer recommendation is relatively new not only to the area of MOOC but also to the recommendation systems and gains more ground with many such applications.


\section{Proposed Method}

In the next few subsections, we talk about our data model along with the design and description of our recommendation algorithm.

\subsection{Data}
The data used in our research comes from the de-identified release from the first year (Academic Year 2013: Fall 2012, Spring 2013, and Summer 2013) of MITx and HarvardX courses on the edX platform \cite{MITx}. For our analysis and without loss of generality, we selected records with attributes about age, location, qualification and gender. Moreover, we enhance this information with synthesized data about learners' interests. This information is not available via the mentioned dataset but is potentially useful for recommending learners to other learners.

A brief overview of the dataset attributes can be found in Table \ref{tab:features}. The \textit{user\_id} is a numerical unique identifier for different learners, \textit{age} of the learner is calculated using the year of birth obtained from the original dataset, \textit{gender} is another binary attribute followed by \textit{location}, which has information about the resident city of the learner. Furthermore, the \textit{qualification} attribute has been divided into 5 levels: {less than secondary, secondary, bachelors, masters and doctorate}. The \textit{interest} attribute contains one or more values about learners' interest. A sample of our dataset can be seen in Table \ref{tab:dataex}.

\begin{table}[!h]
  \vspace{-15pt}
\small
  \centering
  \caption{Dataset Attribute Description}
  \scalebox{0.8}{
    \begin{tabular}{ccccc}
    \hline
    \textbf{Attribute} & \textbf{Short} & \textbf{Type} & \textbf{Comment} \\
   \hline
    user\_id & id    & Numeric & Unique identifier \\
    age   & age   & Numeric & Calculated using year of birth \\
    gender & gen   & Binary & M(ale)/F(emale) \\
    location & loc   & Categorical & City of the learner \\
    qualification & qua   & Ordinal  & 5 levels  \\
    interests & int   & Hierarchical, Categorical, Multi-Value & Info about learners' interests \\
    \hline
    \end{tabular}}
  \label{tab:features}%
      \vspace{-50pt}
\end{table}%

\begin{table}[!h]
\small
  \centering
  \caption{Dataset Sample}
  \scalebox{0.8}{
    \begin{tabular}{ccccccc}
    \hline
    \textbf{id} & \textbf{age} & \textbf{gen} & \textbf{loc} & \textbf{qua} & \textbf{int} & \textbf{crs} \\
    \hline
    1 & 32    & M     & Frankfurt & Doctorate & ML    & machine learning, java, python \\
    2 & 28    & M     & Los Angeles & Bachelors & AI    & java, python \\
    3 & 27    & F     & Edmonton & Bachelors & Science & python, sociology \\
    4 & 22    & F     & Las Vegas & Secondary & Soccer, AI & history, general studies \\
    \hline
    \end{tabular}}
  \label{tab:dataex}%
      \vspace{-30pt}
\end{table}%

\subsection{Preference and Importance Modeling}
When users sign up on a MOOC platform, they provide preferences for the above mentioned attributes, which would be used to recommend similar learners to them. These preferences are based on value ranges for attributes in Table \ref{tab:features}, and can include none, one or more (even all) of these attributes. A description of the value ranges of preferences for each of the attributes is mentioned below:\\\\
-\textbf{Age:} the age preference attribute is divided into these 5 levels: {less than 20, 20-25, 25-30, 30-35, 35 and above}.\\
- \textbf{Gender:} male or female gender options.\\
- \textbf{Location:} \textit{same city} (if learners prefer meeting in person), \textit{same country or timezone} (to facilitate communication).\\
- \textbf{Qualification:} one or more qualifications out of the five levels available.\\
- \textbf{Interests:} users can define their own interest preference which might or might not be similar to their own interest.\\

A sample of user preferences can be seen in Table \ref{tab:pref}. It must be noted that not all five preference attributes are required to be defined by a user. One or more (but not all) of these attributes can be left empty, at which point the algorithm simply ignores these in the recommendation process as it considers them irrelevant. In Table \ref{tab:pref}, `x' denotes no preference for the given attribute by the user.

Moreover, users can further define whether they have some preferences that are more important to them i.e. if they have a priority for their preferences (highlighted in bold in the preference Table \ref{tab:pref}). For instance, looking at preference \textit{p\_4} in Table \ref{tab:pref}, we can tell that this user prioritizes \textit{location} and \textit{qualification} over other attribute preferences.

\begin{table}[!h]
  \vspace{-25pt}
  \centering
  \caption{Sample of User Preferences}
  \scalebox{0.8}{
    \begin{tabular}{cccccc}
    \hline
    \textbf{pref} & \textbf{age} & \textbf{gen} & \textbf{loc} & \textbf{qua} & \textbf{int} \\
    \hline
    p\_1  & \textbf{30-35} & \textbf{M}     & same city & $>=$ Masters & x \\
    p\_2  & x     & x     & x     & Bachelors & Football \\
    p\_3  & \textbf{25-30} & {F}     & x & x     & x \\
    p\_4  & $<=$25   & x     & \textbf{same timezone} & \textbf{$<=$Bachelors} & x \\
    \hline
    \end{tabular}}
  \label{tab:pref}%
    \vspace{-25pt}
\end{table}%

\subsection{Recommendation Algorithm Description}

In the next subsections, we will discuss our recommendation algorithm in detail. In short, we first build a similarity matrix which has the compatibility scores (based on user preferences) between the users. The compatibility scores helps us to generate ranked recommendations. Next, we re-rank the users based on their preference priority.

\subsubsection{Building Similarity Matrix}
Given the preferences of a user, we compute the `distance' of this user, with every other user based on their attribute values. It is to be noted that, `the lower the distance score, the greater the similarity'. For instance, using the data sample in Table \ref{tab:dataex}, distance of a user (with $id$=1) to other users could be computed as follows:\\

- \textbf{Ordinal Variables} (age, qualification): Preferences for \textit{age} and \textit{qualification} attributes are divided into levels in such a way that adjacent levels have a distance of 1, as shown in the data section (3.1).  Once the distance between users for these attributes are calculated, it is then normalized in the range $[0-1]$ by dividing it by the maximum distance possible.

- \textbf{Nominal Variables} (gender, location): Preferences for \textit{gender} and \textit{location} attributes are mapped to a binary distance metric. For instance, if the \textit{gender} of two users are same, then the distance $\textit{`d'}_{gen}$ is 0, otherwise 1. Similarly, the same computation is applied to the \textit{location} or any other nominal variable.

- \textbf{Hierarchical Variables} (interests): For preference attributes that come from a hierarchy there is a similarity measure based on the hierarchy tree. 
This measure, based on the edge counting between nodes by the shortest way, presents a method to evaluate the semantic similarity in a hierarchical tree structure. The hierarchy we used for \textit{interests} of users is based on WordNet \cite{miller1995wordnet} and the similarity measure used is based on the Wu and Palmer method \cite{wu1994verbs} $score$ which considers the depths of the two synsets in the WordNet taxonomies, along with the depth of the LCS (Least Common Subsumer). Score for this similarity is between 0 and 1, since we are implementing our system in a distance measure (and not similarity) the final value of distance between the interests is $[1-score]$.\\

Finally, the `distance score' of a user $x$ with any other user $y$ is the mean of the attribute distances:

\begin{equation}
    distance\_score(x,y)=\frac{\sum_{i=1}^N d_{i}(x,y)}{N}
\end{equation}
where $d_{i}$ is the distance for attribute $i$ between users $x$ and $y$ and $N$ represents the total number of attributes (in our case $N=5$). For instance, the `distance score' of user 3 (\textit{id=3}) with the other users can be computed as follows:\\

$distance\_score(user id:3,user id:1)=\frac{1/4+1}{5}=0.25$, (as $age$ range difference is 1 and $gen$ difference is 1)

$distance\_score(user id:3,user id:2)=\frac{0+1}{5}=0.2$, (as $age$ range is same, but $gen$ is different)

$distance\_score(user id:3,user id:4)=\frac{1/4+0}{5}=0.05$, (as $age$ range difference is 1 and $gen$ is same)

Table \ref{tab:distances} below shows the `Similarity Matrix' with distance scores between all users in the sample dataset (Table \ref{tab:dataex})

\begin{table}[htbp]
    \vspace{-25pt}
  \centering
  \caption{Similarity Matrix}
  \scalebox{1.0}{
    \begin{tabular}{c|cccc}
    \hline
        user\_id  & 1     & 2     & 3     & 4 \\
    \hline
    1     & x     & 0.3   & 0.5   & 0.6 \\
    2     & 0.2   & x     & 0 & 0.15 \\
    3     & 0.25  & 0.2   & x     & 0.05 \\
    4     & 0.45  & 0.1   & 0.3   & x \\
    \hline
    \end{tabular}}
  \label{tab:distances}%
      \vspace{-25pt}
\end{table}%

\subsubsection{Ranking Recommendations by Importance}
After the user preferences and the distance scores are computed, the list of recommended users generated for user $x$ are as follows: Every user $y$ will receive a distance score that reflects how many preferences of user $x$ match with the attributes of user $y$ and vice-versa. We call this measure `reciprocal score'. The reciprocal score between users $x$ and $y$ is the harmonic mean of the distance scores between them. It is to be noted that distance scores of zero are replaced by a small value like $0.001$ in order for the harmonic mean to be computed. A ranking is generated using the reciprocal scores (harmonic mean), it is then verified if the preference priority for attributes as denoted by the user is satisfied or not.\\

For instance, the reciprocal score for user \textit{id:3} is shown in Table \ref{tab:recdist}. Note that the reciprocal score is symmetric as the name suggests, i.e. \textit{y's} score in the recommendation list for \textit{x} is the same as \textit{x’s} score in the list for \textit{y}. However, as the lists contains only the top-N recommendations, user \textit{y} may be in the top-N recommendations for user \textit{x} but the opposite may not be true.

\begin{table}[htbp]
  \centering
      \vspace{-25pt}
  \caption{Reciprocal Score for user \textit{id:3}}
  \scalebox{0.8}{
    \begin{tabular}{c|cc|c}
    \hline
    \textbf{y} & \textbf{p(3,y)} & \textbf{p(y,3)} & \textbf{harmonic\_mean} \\
    \hline
    \textbf{1} & 0.25  & 0.5   & 0.333 \\
    \textbf{2} & 0.2   & 0.001 & \textbf{0.002} \\
    \textbf{4} & 0.05  & 0.3   & 0.086 \\
    \hline
    \end{tabular}}
  \label{tab:recdist}%
      \vspace{-20pt}
\end{table}%

\medskip
Given the reciprocal scores in Table \ref{tab:recdist}, the list of top-3 recommendations for user \textit{id:3} will be: [2, 4 and 1]. Furthermore, user \textit{id:3} has noted preference priority for \textit{age} attribute (see bold values in Table \ref{tab:pref}). Since user \textit{id:2} satisfies this criterion, it will remain at the first position and users \textit{id:4} and \textit{id:1} will follow. If this was not the case, then a re-ranking of recommended users is done based on the preference priority of the user for the given attributes.

\section{Experiments and Results}
\subsection{Evaluation metrics}
The goal of the current work was to primarily explore the role of reciprocity in the formulation of the recommendations for MOOC. It should be noted here that an actual evaluation of a (reciprocal) recommender system requires on-line deployment of the algorithm to one of the existing MOOC platforms. Since this was not possible in our case, we had to build measures based on the data available.

For a reciprocal system (like the one in our case) we need to define `what is a successful recommendation?'. We say that, ``learner $y$ is a successful (reciprocal) recommendation (out of the $K$-total) for learner $x$, if and only if $x$ is also in the top-$K$ recommendations of learner $y$''. This condition factors the reciprocity element which is essential to measure the performance of a reciprocal system like ours. Using this logic, we modify the definitions of precision and recall \cite{shani2011evaluating} for each learner as follows: ``In order to compute the precision for learner $x$, we divide the number of successful recommendations by the total number of recommendations (i.e. $K$) generated for leaner $x$''. ``Similarly, in order to compute the recall for learner $x$, we divide the number of successful recommendations by the total number of learners that have $x$ in their top-$K$ recommendation list''. These definitions can be formalized in the following equations:

\begin{eqnarray}
    P_x = \frac{N_x}{K}, R_x = \frac{N_x}{N*_x}
\end{eqnarray}

where $P_x$ is the precision for learner $x$, $R_x$ is the recall for learner $x$, $N_x$ is the number of successful recommendations for learner $x$ (as defined before), $K$ is the total number of recommendations generated and $N*_x$ is the number of learners that have $x$ in their recommendation list. 

The total precision and recall of the dataset based on the recommendation algorithm is defined as follows:

\begin{eqnarray}
P = \sum_{i=1}^M\frac{P_i}{M}, R = \sum_{i=1}^M\frac{R_i}{M}
\label{eqn:PR}
\end{eqnarray}

where $P_i$ and $R_i$ are the precision and recall respectively for learner \textit{i} (as declared previously) and $M$ is the total number of learners.\\

Moreover, in order to evaluate the rankings of the algorithm, we utilize a modified definition of the Discounted Cumulative Gain (DCG) \cite{jarvelin2002cumulated}, a popular measure of ranking quality. DCG originates from information retrieval where ranking positions are discounted logarithmically. Since for our system, we only care about the rank alignments and not the relevance of ranking positions, hence we do not require the logarithm discounting. When applied to our case, `DCG' is the measure of `reciprocity' or `rank alignment'. In other words, a perfect rank alignment is when - ``for all learners $i$, present at a position $j$ in the list of top-N recommendations of learner $u$, if $u$ is also present at the same position $j$ in the list of top-N recommendations of $i$''.

Assuming each learner $u$ has a ``gain'', $g_{ui}$ from being recommended to another learner $i$, then the average Discounted Cumulative Gain (DCG) for the recommendation list of $K$ learners is defined as follows:

\begin{equation}
DCG = \frac{1}{M} \sum_{u=1}^{M} \frac{\sum_{j=1}^{K}{g_{ui_j}}}{S}
\end{equation}

where $M$ is total number of learners, $S$ is the number of successful recommendations, $j$ denotes the position in the ranking list and $g_{ui_j}$ is the gain of learner $i$ (in position $j$) for learner $u$.

Division by the number of successful recommendations guarantees that maximum DCG will be 1, provided that a user has successful recommendations, otherwise the value is 0.

The gain $g_{ui}$, is 0 if learner $u$ is not in the top-$K$ recommendation list for learner $i$ (no gain for the reciprocal recommendation system here) and if is present, then the gain is defined as follows:

\begin{equation}
g_{ui}=\frac{1}{1+|diff_{ui}|}
\end{equation}

where $diff_{ui}$ is the difference in positions between the ranking of user $i$ in the recommendation list of user $u$ and the ranking of user $u$ in the recommendation list of user $i$. This equation provides a value of 1 if the reciprocal rankings between learners $i$ and $u$ agree, otherwise it discounts this gain.

Finally, DCG can be divided by the ideal DCG for the recommender system which would lead to the normalized discounted cumulative gain (NDCG). Ideal DCG is 1 provided that all users have at least one successful recommendation (each user can have a maximum DCG of 1, so divided by the number of users that gives 1), otherwise it is a reduced value.

\begin{equation}
    NDCG = \frac{DCG}{DCG*}
    \label{eqn:dcg}
\end{equation}

Consider the following Table \ref{tab:measures} of six learners: [1, 2, 3, 4, 5, 6] with successful recommendations highlighted in circles.
\begin{table}[H]
    \vspace{-25pt}
  \centering
  \caption{Ranked Recommendations, $K$=3}
  \scalebox{0.8}{
\begin{tabular}{rcccccc}
\hline
\multicolumn{1}{c}{rank/learner} & {1} & {2} & {3} & {4} & {5} & {6} \\
\hline
\multicolumn{1}{c}{1} & 2     & \textit{\textcircled{3}} & \textit{\textcircled{1}} & 6     & 1     & 3 \\
\multicolumn{1}{c}{2} & \textit{\textcircled{3}} & 4     & \textit{\textcircled{2}} & \textit{\textcircled{5}} & \textit{\textcircled{4}} & 2 \\
\multicolumn{1}{c}{3} & 4     & 5     & \textit{\textcircled{4}} & \textit{\textcircled{3}} & 6     & 1 \\
\multicolumn{1}{c}{4} & 5     & 1     & 5     & 1     & 2     & 4 \\
\multicolumn{1}{c}{5} & 6     & 6     & 6     & 2     & 3     & 5 \\
\hline
Precision & 0.33  & 0.33  & 1.00  & 0.67  & 0.33  & 0.00 \\
Recall & 0.33  & 0.33  & 0.75  & 0.50  & 0.50  & 0.00 \\
DCG   & 0.50 & 0.50 & 0.67 & 1.00 & 1.00 & 0.00 \\
\hline
\end{tabular}}
  \label{tab:measures}%
      \vspace{-20pt}
\end{table}
Overall precision for this system is 0.44, recall is 0.40 and the NDCG is 0.73 (DCG is 0.61 and DCG* is 0.83).

\medskip
We conducted our experiments with 5 different samples of 1000 user records from the dataset. From each of these samples, we ranked users by comparing their reciprocal scores and recommended the top-N [5,10,15,20] users in the list. The results were averaged across the samples for each of these top-N recommendations. Our precision, recall and `DCG' scores are compared against the `baseline', wherein the reciprocity factor was not accounted for. The `baseline' model builds the list of top-N recommendations without looking at reciprocity, very similar to a traditional recommender system.

\begin{figure}[!h]
    \vspace{-20pt}
\begin{subfigure}
    \centering
    \includegraphics[width=0.55\textwidth]{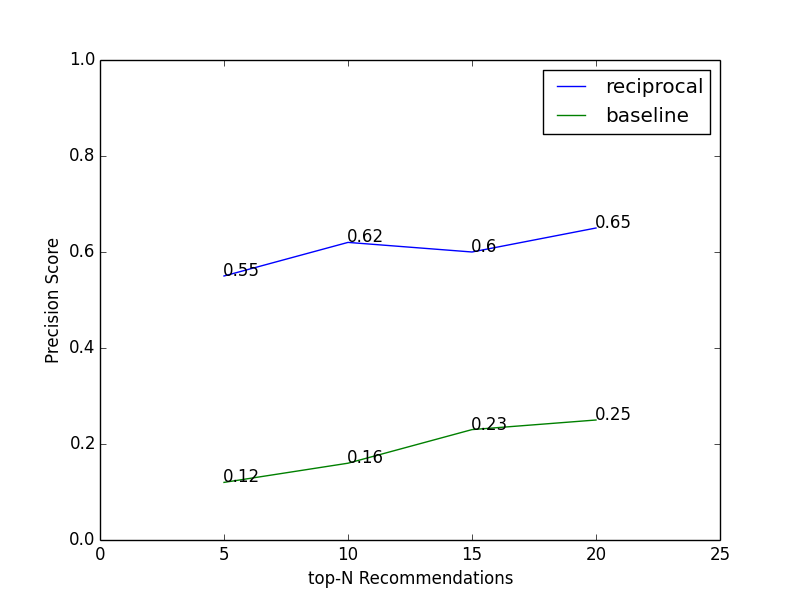}
\end{subfigure}
\begin{subfigure}
    \centering
    \includegraphics[width=0.55\textwidth]{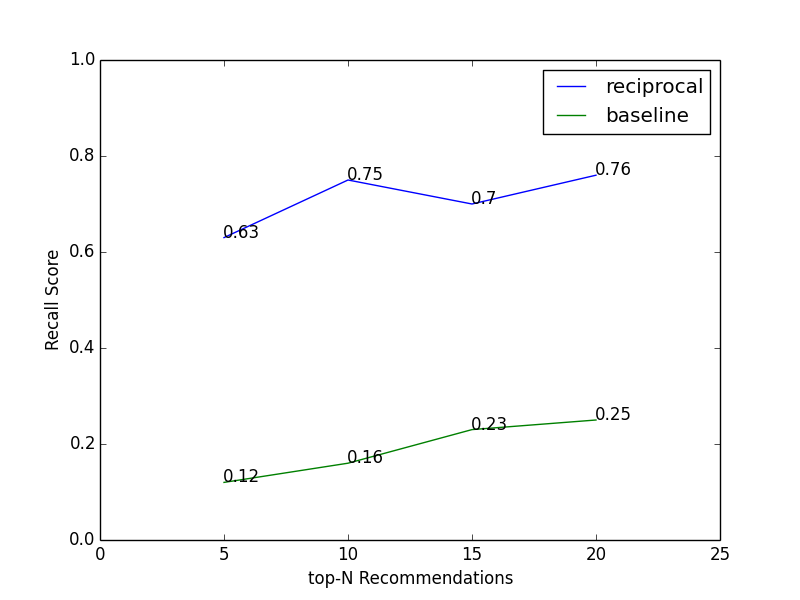}
\end{subfigure}
\caption{Precision and Recall Graph}
\label{fig:p_nc}
    \vspace{-25pt}
\end{figure}

The precision and recall graphs are shown in Figure \ref{fig:p_nc}. As expected, precision and recall increase with `N', which means that in the case of precision, if a learner $y$ is present in the top-N recommendation list for learner $x$, then the chances that $x$ is also present in the recommendation list of $y$ increases with increasing value of `N'. The same holds true for recall as well.

\begin{figure}[!h]
    \vspace{-20pt}
\centering
\includegraphics[width=0.55\textwidth]{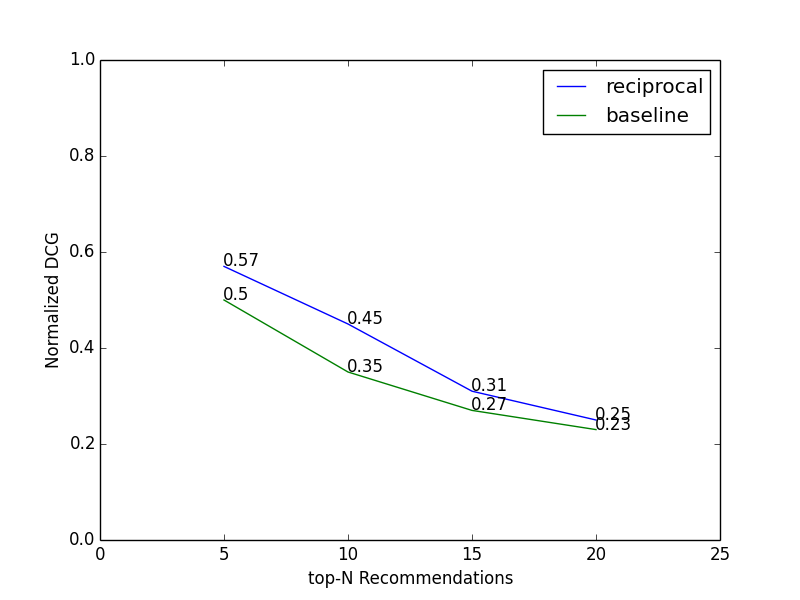}
\caption{NDCG Graph}
\label{fig:ndcg_nc}
\centering
    \vspace{-25pt}
\end{figure}

We also calculate the `Normalized DCG' or `NDCG' as shown in Figure \ref{fig:ndcg_nc}. The value of `NDCG' decreases if the `top-N' recommendations increase. This makes sense because with higher number of recommendations, the difference in ranks for two positions in the recommendation list will increase, thereby resulting in an overall decrease in `gain'.\\

In summary, the precision and recall scores for `reciprocal' model far exceeds the scores for `baseline' model whereas the `NDCG' values for `reciprocal' is slightly better than the baseline model across all values of top-N recommendations.

\section{Conclusions and Future Work}
In this paper we proposed an algorithm that allows learners to reach out and communicate with other similar learners and it thereby facilitates meaningful discussions and encourages peer learning. Results show that our system performs better than the baseline system on the measures of precision, recall and discounted cumulative gain. As future work, we plan to incorporate some more learner attributes like `communication frequency', `leadership ability' etc., based on the historical interaction of users on various MOOC forums. This will certainly help to improve the list of recommendations. Moreover, we plan to conduct tests on an actual MOOC platform to measure the quality of recommendations. Such an experimentation will evaluate how reciprocal recommendation can improve learners' experience but it should be noted that it requires longer times and greater difficulties in implementation. Case studies reveal that with the number of participating users in MOOCs increasing exponentially every year, it is quite challenging to establish the same kind of communication that exists within a classroom. However, with this proposed model, we believe we can bridge that gap to some extent.

\bibliographystyle{plain}
\bibliography{bibliography}

\end{document}